\documentclass[twocolumn]{aastex63}

\usepackage{graphicx}
\usepackage{amsmath}
\usepackage{amssymb}
\usepackage{txfonts}

\usepackage{ulem}

\usepackage{color}
\usepackage{colortbl} 
\definecolor{pinegreen}{RGB}{1, 121, 111}
\definecolor{salmon}{RGB}{255,160,122}
\usepackage{natbib}

\definecolor{YC}{RGB}{115,80,185}

\newcommand{\Msun}{\ensuremath{\,M_\odot}}
\newcommand{\Rsun}{\ensuremath{\,R_\odot}}

\newcommand{\kms}{\ensuremath{\,\rm{km}\,\rm{s}^{-1}}}
\newcommand{\Msunyr}{\ensuremath{\,M_\odot\,\rm{yr}^{-1}}}
\newcommand{\ergs}{\ensuremath{\,\rm{erg}\,\rm{s}^{-1}}}

\newcommand{\CD}{CD$-30^{\circ}$}
\newcommand{\ZJ}{ZTF~J$2130$}

\usepackage{refs}   
\usepackage{booktabs}    
\usepackage{morefloats}

\newcommand{\degree}{\ensuremath{{}^{\circ}}}

\bibpunct{(}{)}{;}{a}{}{,}

\submitjournal{ApJ}

\begin{document}

\shorttitle{Stripped stars as living gravitational wave sources}
\shortauthors{G\"{o}tberg et al.}

\title{Stars stripped in binaries -- the living gravitational wave sources}

\correspondingauthor{Y.~G\"{o}tberg}
\email{ygoetberg@carnegiescience.edu}

\author[0000-0002-6960-6911]{Y.~G\"{o}tberg}
\affiliation{The observatories of the Carnegie institution for science, 813 Santa Barbara Street, Pasadena, CA 91101, USA}
\affiliation{Kavli Institute for Theoretical Physics, University of California, Santa Barbara, CA 93106, USA}

\author[0000-0002-6725-5935]{V.~Korol}
\affiliation{School of Physics and Astronomy \& Institute for Gravitational Wave Astronomy, University of Birmingham, Birmingham, B15 2TT, UK}

\author[0000-0001-8740-0127]{A.~Lamberts}
\affiliation{Universit\'e C\^ote d'Azur, Observatoire de la C\^ote d'Azur, CNRS, Laboratoire Lagrange, Laboratoire ARTEMIS, France}
\affiliation{Kavli Institute for Theoretical Physics, University of California, Santa Barbara, CA 93106, USA}

\author[0000-0002-6540-1484]{T.~Kupfer}
\affiliation{Kavli Institute for Theoretical Physics, University of California, Santa Barbara, CA 93106, USA}

\author[0000-0001-5228-6598]{K.~Breivik}
\affiliation{Canadian Institute for Theoretical Astrophysics, University of Toronto, 60 St.  George Street, Toronto, Ontario, M5S 1A7, Canada}

\author[0000-0003-0857-2989]{B.~Ludwig}
\affiliation{Department of Astronomy and Astrophysics, University of Toronto, 50 St. George Street, Toronto, Ontario, M5S 3H4, Canada}

\author[0000-0001-7081-0082]{M.~R.~Drout}
\affiliation{Department of Astronomy and Astrophysics, University of Toronto, 50 St. George Street, Toronto, Ontario, M5S 3H4, Canada}

\begin{abstract}
Binary interaction can cause stellar envelopes to be stripped, which significantly reduces the radius of the star. The orbit of a binary composed of a stripped star and a compact object can therefore be so tight that the gravitational radiation the system produces reaches frequencies accessible to the Laser Interferometer Space Antenna (LISA). Two such stripped stars in tight orbits with white dwarfs are known so far (ZTF J2130+4420 and CD$-30^{\circ}11223$), but many more are expected to exist. These binaries provide important constraints for binary evolution models and may be used as LISA verification sources.
We develop a Monte Carlo code that uses detailed evolutionary models to simulate the Galactic population of stripped stars in tight orbits with either neutron star or white dwarf companions. 
We predict $0-100$ stripped star + white dwarf binaries and $0-4$ stripped star + neutron star binaries with SNR $>5$ after 10 years of observations with LISA. More than $90$\% of these binaries are expected to show large radial velocity shifts of $\gtrsim 200 \kms$, which are spectroscopically detectable. Photometric variability due to tidal deformation of the stripped star is also expected and has been observed in ZTF J2130+4420 and CD$-30^{\circ}11223$. In addition, the stripped star + neutron star binaries are expected to be X-ray bright with $L_X \gtrsim 10^{33} - 10^{36} \ergs$. Our results show that stripped star binaries are promising multi-messenger sources for the upcoming electromagnetic and gravitational wave facilities. 
\end{abstract}

\keywords{Compact binary stars (283), Interacting binary stars (801), Common envelope binary stars (2156), Gravitational wave sources (677), Radial velocity (1332), Ellipsoidal variable stars (455), X-ray binary stars (1811)}

\section{Introduction}

The recent detections of gravitational waves (GWs) from merging binary black holes (BHs) and binary neutron stars (NSs) have placed double compact objects in the spotlight \citep{2016PhRvL.116v1101A, 2017PhRvL.119p1101A}. However, GW sources are not limited to binary systems composed of stellar remnants; GWs can also be generated by binaries containing ``living'' stars -- stars still undergoing nuclear burning in their interior. 

Most living stars are too large to fit in the tight orbits needed for the emitted gravitational radiation to be detectable. But some living stars are compact and small: stars that have been stripped of their hydrogen-envelopes. Envelope-stripping can occur in binary stars either via stable mass-transfer or through the successful ejection of a common envelope \citep[e.g.,][]{1967ZA.....65..251K, 1967AcA....17..355P, 2011ApJ...730...76I}. This process is predicted to be the fate for $\sim 30\%$ of all massive stars \citep{2012Sci...337..444S} and to primarily occur prior to the central helium burning phase, which constitutes about 10\% of the total stellar lifetime. The resulting stripped stars are hot ($\gtrsim 30$kK) and helium-rich. Low-mass stripped stars are classified as subdwarfs, since they have absorption line spectra. With increasing mass, the stellar winds are expected to be stronger, suggesting that high-mass stripped stars can have emission line spectra and be classified as Wolf-Rayet (WR) stars \citep{2018A&A...615A..78G}. As envelope-stripping is predicted to be common and the remaining lifetime is significant, stripped stars should be relatively abundant in the Milky Way \citep{2019A&A...629A.134G}. 

\begin{figure}
\centering
\includegraphics[width=\columnwidth,trim=90mm 15mm 100mm 20mm,clip]{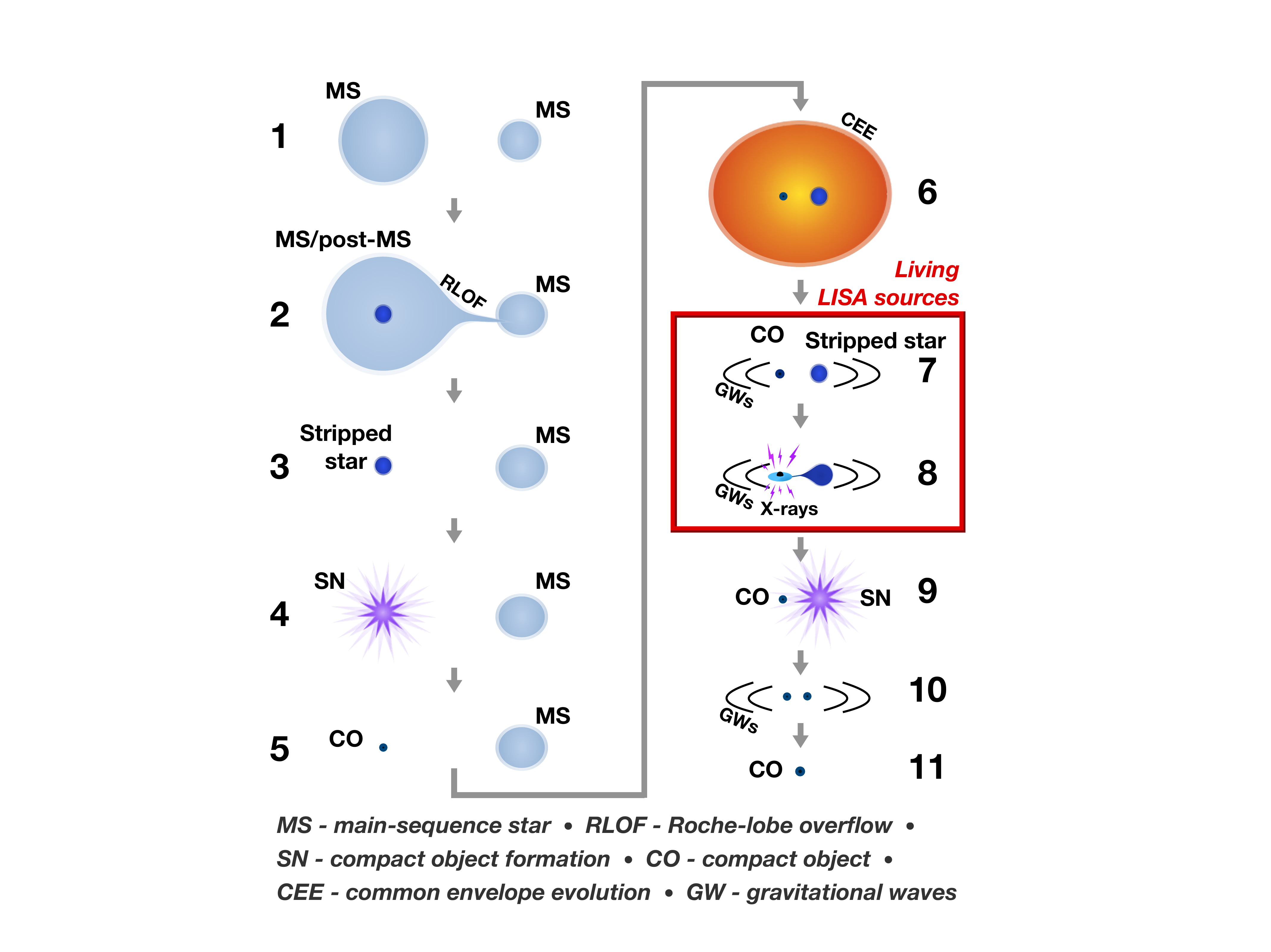}
\caption{Visualization of a common evolutionary pathway for the formation of a stripped star in tight orbit with a compact object. The red box marks when a stripped star orbits a compact object close enough that the binary could be detectable by LISA. Each evolutionary phase is numbered according to their order in the evolution and the direction of the sequence is marked with arrows. In stage 2 and 6 the two stars are relieved of their hydrogen-rich envelopes, creating a stripped star. The envelope-stripping in stage 2 is shown as Roche-lobe overflow, but the envelope can also be lost via the ejection of a common envelope during this stage. We show the compact object formations at stage 4 and 9 with a purple star and marked with `SN', but note that when white dwarfs are formed they do not experience an explosion.
At stage 8, we indicate X-ray emission from accretion onto the compact object via Roche-lobe overflow. We note that the X-ray emission from most systems is expected to originate from wind accretion, meaning that X-ray emission is also expected for systems in stage 7 (see \secref{sec:Xrays}).
}
\label{fig:cartoon}
\end{figure}

Stripped star binaries can also be the direct progenitors of double compact objects. This is illustrated in \figref{fig:cartoon}, which shows the evolutionary sequence that is thought to result in most of the mergers of double NSs or double BHs \citep[e.g.,][see however also \citealt{2020arXiv200109829V}]{2017ApJ...846..170T, 2017NatCo...814906S}. In particular, the figure shows that envelope-stripping occurs twice before the merger of the two compact objects (step 2 and 6), acting as a key ingredient in bringing the stars close together \citep[e.g.,][]{2003ApJ...592..475I}. In the case of merging double white dwarfs (WDs), both of the stellar envelopes are likely ejected through common envelope phases \citep[e.g.,][]{2012A&A...546A..70T}.

When stripped stars are created through common envelope evolution (step 6-7, \figref{fig:cartoon}), the resulting binary has a short orbital period \citep{2016MNRAS.460.3992N, 2019ApJ...883L..45F}. If, in addition, the companion star is a compact object, both stars are so small that the orbit can be sufficiently short for GWs in the sub-mHz regime to be emitted.
This means that their emitted GWs are not detectable by the \textit{Laser Interferometer Gravitational-Wave Observatory} (LIGO) \citep{2015CQGra..32g4001L} or the \textit{Virgo interferometer} \citep{2015CQGra..32b4001A}, which operate in the $\sim 10^2- 10^4$ Hz range. Rather, if the signal is sufficiently strong, they would appear in the band of the \textit{Laser Interferometer Space Antenna} \citep[LISA,][]{2017arXiv170200786A}. With the decreasing sensitivity of LISA at the low-frequency end, these systems are ideally detected by $\mu$Hz GW instruments, such as the recently proposed $\mu$-ARES mission \citep{2019arXiv190811391S}.

In contrast to many other gravitational wave sources, stripped stars in tight orbit with compact objects are bright in the electromagnetic spectrum, which makes them possible to study in advance of the launch of LISA. These systems will be valuable verification sources for LISA, allowing calibration of the instrument at frequencies lower that the those covered by double WDs \citep[e.g.][]{2018MNRAS.480..302K}. 

Despite their predicted existence and crucial role in the creation of double NSs, no stripped star has been observationally confirmed to orbit a NS \citep[see however the debated system  HD~49798,][]{1997ApJ...474L..53I}. The evolution of such systems have been studied theoretically \citep{2002MNRAS.331.1027D, 2003MNRAS.344..629D, 2018A&A...618A..14W} leading to predictions for the population of low-mass stripped stars orbiting NSs \citep{2020A&A...634A.126W}. 
Validating the existence of stripped stars in orbit with NSs and studying their properties is of highest importance for verifying binary evolution models and for our understanding of the formation of merging compact objects. Population properties, such as the total number of systems, and their masses and orbital periods can constrain, for example, the outcome of the common envelope evolution, which is considered one of the most uncertain binary evolution processes \citep[e.g.,][]{2003ApJ...592..475I}. 
Detecting the population of stripped star binaries that emit GWs will therefore provide valuable constraints.

LISA itself will also help to identify stripped star systems since dust extinction is a problem for electromagnetic searches but does not affect the detectability of GWs. Therefore, combining electromagnetic searches with the detections of LISA will give a more complete picture of the stripped star binary population. 

Two low-mass stripped star systems containing WDs are already found to have orbital parameters detectable with LISA. 
CD$-30\degree11223$ (hereafter \CD) is a 0.5\Msun\ subdwarf in a detached, 70-minute orbit with a 0.7\Msun\ WD \citep{2012ApJ...759L..25V, 2013A&A...554A..54G, 2018MNRAS.480..302K} and 
ZTF J2130+4420 (hereafter \ZJ) is a system containing a 0.3\Msun\ subdwarf that transfers material to its 0.5\Msun\ WD companion. The orbital period of \ZJ\ is  short compared to \CD, only 39 minutes \citep{2020ApJ...891...45K}. 
Systems like \CD\ with a high WD mass are especially interesting as progenitors for thermonuclear supernovae through the sub-Chandrasekhar scenario. In this scenario, the stripped star is thought to initiate mass transfer after the helium core burning phase. After the WD has accreted $\sim 0.1\Msun$, helium is predicted to be ignited in a shell on the surface of the WD. This in turn triggers the ignition of carbon in the core even if the WD mass is significantly lower than the Chandrasekhar limit \citep{2010A&A...514A..53F}. So far, \CD\ is the only known candidate for this scenario. 

In this paper, we model the Galactic population of stripped stars in tight orbits with compact objects, focusing on those that will be detectable by LISA. Throughout this manuscript, we refer to these systems as stripped star binaries, but note that stripped stars are expected to most often be accompanied by main-sequence stars. 
By considering a wide mass range for stripped stars and accounting for both white dwarf and neutron star companions, we provide first predictions for a type of gravitational wave source that previously has not been considered. 
Our study aims at exploring the parameter space that LISA will be sensitive to (in terms of stellar properties and numbers of the stripped star binaries). 
In addition to electromagnetic detections, confirmed GW signals will provide independent constraints on the population of stripped stars in tight binaries with compact objects. 
Apart from providing important constraints on binary evolution, observations of stripped stars orbiting compact objects will also allow to investigate the effect of gravitational waves on stellar interiors.  
We, therefore, also discuss a number of promising methods for identifying these binaries in advance of the LISA launch planned in the early 2030's.

We structure the article as follows. In \secref{sec:modeling}, we describe how we combine detailed models of stripped stars with simple assumptions for a population, including their mass, age, and spatial distributions in the Milky Way. We also calculate the orbital tightening due to GW radiation, which impacts whether binary interaction is initiated anew (step 7 to 8 in \figref{fig:cartoon}). In \secref{sec:properties}, we present the predicted properties of the binary systems and estimate their GW signal. In \secref{sec:EM}, we discuss electromagnetic detection techniques and, finally, in \secref{sec:summary}, we summarize our findings. 


\section{Modeling a population}\label{sec:modeling}

Our goal is to simulate the Galactic population of stripped stars in tight orbits with compact objects and to characterize if their gravitational wave signals will be detectable by LISA. We take a Monte Carlo approach to model the population of these systems. 
As a baseline model, we assume that all such systems follow a similar evolutionary pathway as sketched in \figref{fig:cartoon}. This means that the initially most massive star in the system loses its envelope either via mass transfer or common envelope evolution, leaving a stripped star orbiting a main-sequence star (steps 2-3). Later, when this first stripped star evolves to either a white dwarf or neutron star, the system remains bound (steps 4-5). Subsequently, the companion star evolves and swells up and when interaction is initiated, the mass transfer is unstable, which leads to the development of a common envelope (step 6). After the successful ejection of the common envelope, a tight-orbit binary composed of a stripped star and a compact object is created (step 7). This second stripped star may fill its Roche lobe either as a result of orbital tightening from gravitational wave emission or because it evolves and swells up (step 8). 

Here, we describe our initial set-up, the assumptions we make to evolve the systems, and the star formation rate and spatial distribution of stars that we use to model the Milky Way. This section describes our ``standard model'', while the impacts of adjusting various assumptions on the number and properties of stripped star systems detectable by LISA are described in \secref{sec:variations}. 

\subsection{Initial Model Set-up}

As a starting point for our simulation, we take the mass-dependent number density of stripped stars that are created during the first interaction phase (step 2-3 in \figref{fig:cartoon}) calculated by \citet{2019A&A...629A.134G}. For this, \citet{2019A&A...629A.134G} started with the initial mass function from \citet{2001MNRAS.322..231K}, the mass-dependent binary fraction presented in \citet{2017ApJS..230...15M}, the period distributions from \citet{1924PTarO..25f...1O} and \citet{2012Sci...337..444S}, and a uniform distribution in mass ratio \citep[see e.g., ][]{2012ApJ...751....4K, 2012Sci...337..444S}. They then coupled these distributions with the detailed binary evolution models of \citet{2018A&A...615A..78G} to calculate the number of stripped stars created from stars with initial masses between 2 and 20\Msun\ in a stellar population as a function of time.

Metallicity is thought to affect the mass, radius, and future evolution of stripped stars \citep{2017ApJ...840...10Y, 2017A&A...608A..11G, 2019ApJ...885..130S, 2019MNRAS.486.4451G, 2020arXiv200301120L}, which could alter the number of stripped star binaries and their appearance. However, the strongest effects are predicted to occur at very low metallicities, which is not expected to be common in the Galactic thin disk \citep{2014A&A...562A..71B}, where most of the stripped star binaries should reside since they are young stars. 
In simulating the stripped star population of the Milky Way, we therefore assume a single composition with solar metallicity \citep{2009ARA&A..47..481A}.

Following \citet{2019A&A...629A.134G}, we model systems in which the most massive star initially was between 2 and 20\Msun. 
This mass range excludes subdwarfs formed through low-mass stellar evolution. Low-mass stars ($<2\Msun$) that form subdwarfs are thought to initiate the common envelope phase at the tip of the red giant branch when they have large orbital separations. The resulting binary is therefore also relatively wide \citep{2003MNRAS.341..669H} and thus it is unlikely to be a strong GW emitter. The adopted mass range also excludes the most massive stripped stars, some of which are predicted to be progenitors of the BHs with masses $>30\,$\Msun\ observed by LIGO \citep{2019PhRvX...9c1040A}. Massive stripped star binaries are likely to be only a few in the Galaxy \citep[cf.\ Cygnus X-3,][]{1992Natur.355..703V}, although would be an interesting avenue for future study. 


\subsection{Compact Object Formation}

A compact object is created as a result of the death of the first stripped star, as illustrated in step 4-5 in \figref{fig:cartoon}. Depending on the mass of the stripped star, either a WD or a NS is formed. We assume that stripped stars evolve into 1.4\Msun\ NSs if they had initial masses of $>9$\Msun\ and into WDs if they had initial masses $\leq 9$\Msun. This dividing line translates to a stripped star mass of 2.5\Msun\ \citep{2018A&A...615A..78G}. To estimate masses of the WDs, we use the initial to final mass relation constructed using observations of WDs and single star models from the MIST catalog \citep[see][and references therein]{2018ApJ...866...21C}. Since the relation extends to initial masses of 7.2\Msun, we extrapolate to higher masses to reach at most WD masses of 1.3\Msun. As a star with initially 9\Msun\ corresponds to a stripped star with 2.5\Msun, we thus assume that mass is lost when high-mass WDs are created. The WD mass relation predicts WD masses higher than their stripped star progenitors for masses below $\sim 0.9\Msun$. In this regime, we assume that the WDs have the same mass as their stripped star progenitors.
Which stripped stars leave a BH (and not a NS) after core collapse is uncertain. Here, we consider this outcome to be unlikely for the mass range that is concerned in this study \citep{2020ApJ...890...51E}. However, stripped stars orbiting BHs are expected to exist \citep{2020arXiv200411821Y} and one recently observed star has been considered as a candidate \citep[see][]{2020A&A...633L...5I, 2020arXiv200412882S}.

\subsection{Formation and Properties of a Second Stripped Star}\label{sec:2G_strip}

There are numerous uncertainties associated with the survival of binary systems through both compact object formation (\figref{fig:cartoon}; step 4) and common envelope ejection (\figref{fig:cartoon}; step 6). Together, they make it difficult to accurately predict the number of stripped stars in tight orbits with compact objects (\figref{fig:cartoon}; steps 7-8, see also \citealt{2020A&A...634A.126W} and \citealt{2020arXiv200302480L}). 
Here we assume that the fraction of systems that form a first stripped star after the first interaction phase also form a second stripped star, $f_{\text{eff}}$, is $10\%$. This is an approximate but realistic assumption since many systems are disrupted by the instantaneous mass-loss associated with the creation of the compact object and in other cases the subsequent common envelope results in coalescence. Furthermore, the assumed fraction is also supported at the high-mass end by the results of \citet{2017ApJ...842..125Z}, who used binary population synthesis and found that about $7\%$ of stripped stars that have a companion star at explosion have a compact companion. We note, however, that \citet{2017ApJ...842..125Z} studied stripped envelope supernovae in general and therefore accounted for the full mass range, which could change the fraction.

Once a second stripped star is formed, we next assume that it will have the same mass as the first stripped star created in that binary system (i.e., steps 3 and 7 in \figref{fig:cartoon}). This simplified approximation is supported because the first phase of mass transfer can be conservative, causing the progenitor of the second stripped star to gain mass and thus also allowing it to be more massive \citep[step 2 in \figref{fig:cartoon}, see][]{2015ApJ...805...20S}. An example of a system that went through conservative mass transfer is $\varphi$~Persei \citep{1998ApJ...493..440G, 2018A&A...615A..30S}. Furthermore, the components of binary stars tend to initially often have similar masses \citep{2012Sci...337..444S, 2017ApJS..230...15M}. 

With the mass of the second stripped star established, we estimate their stellar radii by the interpolating the structure models of \citet{2018A&A...615A..78G} over their initial masses. We adopt radii that are measured halfway through central helium burning ($X_{\text{He,c}}=0.5$). During the long-lasting helium core burning phase, stripped stars do not expand or contract significantly \citep[see e.g., Appendix C of][]{2019A&A...629A.134G} and, therefore, we do not account for their radius evolution in our simulations (see however the discussion in \secref{sec:properties}).

We acknowledge that the models of \citet{2018A&A...615A..78G} are constructed for stars stripped of their hydrogen-rich envelopes via stable mass transfer, while here we consider stars that are stripped via the ejections of common envelopes. It is likely that more material is lost during the more violent common envelope evolution compared to stable Roche-lobe overflow, resulting in somewhat lower masses, smaller stars, and less left-over hydrogen on the surface \citep[e.g.,][]{2011ApJ...730...76I, 2017ApJ...840...10Y,2019ApJ...883L..45F}. Nonetheless, after inspecting the solar metallicity stripped star models of \citet{2018A&A...615A..78G}, we found that the remaining mass of hydrogen is very small. Consequently, we expect that it causes the radius to be at most 0.01\Rsun\ larger. This is less than the radius evolution the star experiences during central helium burning. We, therefore, consider that the models of \citet{2018A&A...615A..78G} are appropriate to use to represent the physical properties of stripped stars created via unstable mass transfer.

\subsection{Post-Common Envelope Orbits}

Common envelope evolution is thought to result in very tight, circular orbits if the stars do not coalesce \citep[e.g.,][]{2018A&A...618A..14W, 2019ApJ...883L..45F}. 
Systems that survive a common envelope phase are expected to have typical periods prior to interaction between 100 to 1000 days \citep[e.g.,][]{2008AIPC..990..230D, 2020arXiv200300195V}. Using the $\alpha$-prescription \citep{1984ApJ...277..355W} and assuming a standard efficiency parameter of $\alpha = 1$ and the envelope structure parameter to be $\lambda = 0.5$ \citep{2000A&A...360.1043D,2004PhDT........45I}, we estimate that the orbital periods of the resulting stripped star binaries can be a few to several hours. However, there are many uncertainties associated with the ejection of a common envelope and the resulting period distribution is poorly understood. Studies have suggested both efficient and in-efficient ejection of the common envelope \citep{2010A&A...520A..86Z, 2019ApJ...883L..45F}.

Therefore, for simplicity, in our ``standard model'' population we assume that the orbits after the common envelope ejection are circular and that the orbital period is uniformly distributed between the shortest possible period before interaction starts, $P_{\min}$, and $3 \times \,P_{\min}$. This corresponds to periods between 1 and 10 hours. 
We calculate $P_{\min}$ by setting the Roche radius of the stripped star \citep{1983ApJ...268..368E} equal to the stellar radius (see \secref{sec:2G_strip}).

\subsection{Evolution of the Orbits}

The GW signal from a stripped star binary will be dependent on its orbital separation at the time of observation. We account for the effect that GW radiation has on the evolution of the orbital periods after common envelope evolution following \cite{1964PhRv..136.1224P}. 
We also investigated whether tidal forces would meaningfully impact the orbits following \citet{2019ApJ...885L...2P}. We find that the effect of tides is negligible, mainly because the binary orbits are too wide. Thus, we do not include tides when calculating the orbital evolution in our Monte Carlo simulation. 

In some cases, the orbital tightening causes the stripped star to fill its Roche lobe. If the subsequent mass transfer proceeds in a conservative manner, this can lead to orbital widening. As the widening from mass transfer counteracts the tightening from gravitational waves, systems can stall their orbital evolution at roughly the orbital period at which they initiated mass transfer \citep[e.g.,][]{2017ApJ...846...95K}. To better understand the orbital evolution of systems in the later case, careful evolutionary modeling of the individual systems are needed \citep[see][]{2008AstL...34..620Y, 2017ApJ...847...78B, 2018A&A...618A..14W}. As an approximation, in this work we assume that when the mass transfer is initiated, the orbital evolution stops. We keep these systems in the Galactic population for the rest of their lifetime and we label them as mass transferring.

\subsection{Simulating the Milky Way: Distributions of Stars}

In order to determine the GW signals that LISA will detect we must project our population, simulated in terms of the mass-dependent number density of stripped stars, onto age and spatial distributions consistent with the Milky Way. Since stripped stars and their progenitors are relatively young stars ($\sim 10-1\,000$~Myr, see \citealt{2019A&A...629A.134G}), we assume that their spatial distribution follows the current star-formation in the Milky Way. In particular, \citet{2012ARA&A..50..531K} showed that star-formation is relatively smoothly and uniformly distributed in the thin disk, which has a small typical scale height of $\lesssim0.2\,$kpc \citep{2017MNRAS.471.3057M}. We therefore approximate the thin disk to be infinitesimally thin and also assume that it extends from the Galactic center out to 15~kpc. For our ``standard model'', we randomly distribute stripped star systems in the thin disk, assuming a constant star-formation rate of 2 \Msunyr\ \citep{2011AJ....142..197C,2018ApJS..237...33X}. Then, we calculate the distances to each system by assuming that the Sun is located 8~kpc from the Galactic center \citep[e.g.,][]{2019A&A...625L..10G}.

\section{Properties}\label{sec:properties}

\subsection{Full Population}

We find that the total Galactic population of stripped stars in tight orbit with a compact object is large. Our model predicts that, at the present time, there are about 90\,000 
stripped stars with WD companions and about 5\,000 
stripped stars with NS companions in the Milky Way.
The masses of the stripped stars with WD and NS companions are $0.3-2.5\Msun$ and $2.5-7.4\Msun$, respectively. The chirp masses, defined as $\mathcal{M}_{\text{chirp}} = (m_1m_2)^{3/5}/(m_1+m_2)^{1/5}$, where $m_1$ and $m_2$ are the masses of the two stars in the system, are thus in the ranges $0.3-1.0\Msun$ and $1.6-2.6\Msun$ for stripped stars orbiting WDs and NSs respectively. 
The orbital periods of $\sim 1-10$ hours correspond to GW frequencies in the range $f_{\text{GW}} = 0.06 - 0.5$~mHz. In our model, NSs are assumed to have more massive stripped star companions than WDs \citep[cf.][]{2018A&A...618A..14W} and since massive stripped stars are also larger, their maximum GW frequencies are limited to $\lesssim 0.2$~mHz. 

We find that 15\% of the stripped stars with WD companions have started mass transfer because of the orbital tightening from GW radiation. In contrast, we find that only $0.3\%$ of the stripped stars with NS companions have initiated mass transfer. The reason for this difference is due to the difference in age: lower mass stripped stars can be much older, allowing the GW radiation to act over longer timescales.

The orbital tightening due GW radiation does not induce an observable time derivative of the GW frequency, $\dot{f}_{\text{GW}}$. In our simulated population $\dot{f}_{\text{GW}} \lesssim 2\times 10^{-19}$~Hz$^2$, while for an observation time of $10$~years, the minimum detectable frequency derivative for LISA is $\dot{f}_{\text{GW, det}} = 1/T_{\text{obs}}^2 \approx 10^{-17}$~Hz$^2$. We, therefore, conclude that the stripped star systems will appear as monochromatic GW sources.

\subsection{Detectable by LISA}

We take the approach of \citet{2003PhRvD..67j3001C} to calculate the orbit-averaged gravitational wave amplitudes, $\langle \mathcal{A} \rangle$, for stripped stars in tight orbits with compact objects. For this calculation, we assume that the binaries have a random inclination, $i$, drawn uniformly between $-1$ and $1$ from $\cos i$ probability density function -- this is to guarantee that all viewing angles between the orbital plane and the line of sight are equally likely -- and a random polarization angle drawn from a uniform distribution between 0 and $\pi$. We then calculate the characteristic, dimensionless gravitational wave strain, $h_c$, that the sources would have in LISA following $h_c = \langle \mathcal{A} \rangle \sqrt{f_{\text{GW}} T_{\text{obs}}} $ \citep[e.g.,][]{2018MNRAS.480..302K}. 
We calculate the signal-to-noise ratios (SNR) as $\mathrm{SNR} = \langle \mathcal{A} \rangle \sqrt{T_{\text{obs}}}/S_n (f_{\text{GW}})$, where $S_n (f_{\text{GW}})$ is the square root of LISA's power spectral density at the frequency emitted by the binary. This adopted noise curve corresponds to the mission proposal design \citep{2017arXiv170200786A} and accounts for the confusion noise produced by unresolved Galactic double WDs from the population modeled by \citet{2017MNRAS.470.1894K}. 
In this study, we consider the maximum LISA operation time of $T_{\text{obs}} = 10$ years instead of the conservative nominal duration of 4 years.

Because we take a Monte Carlo approach and the code predicts low numbers, we run the code 1000 times. We then measure the mean value and standard deviations by fitting Poisson distributions to the number of detectable systems, which we here define as systems with SNR~$>5$ \citep{CrowderCornish}. Below, we first describe the results when assuming the previously described set-up. Then, we explore the dependence that the results have on several uncertain parameters by presenting the results when parameters are varied.

\begin{figure*}
\centering
\includegraphics[width=0.75\textwidth]{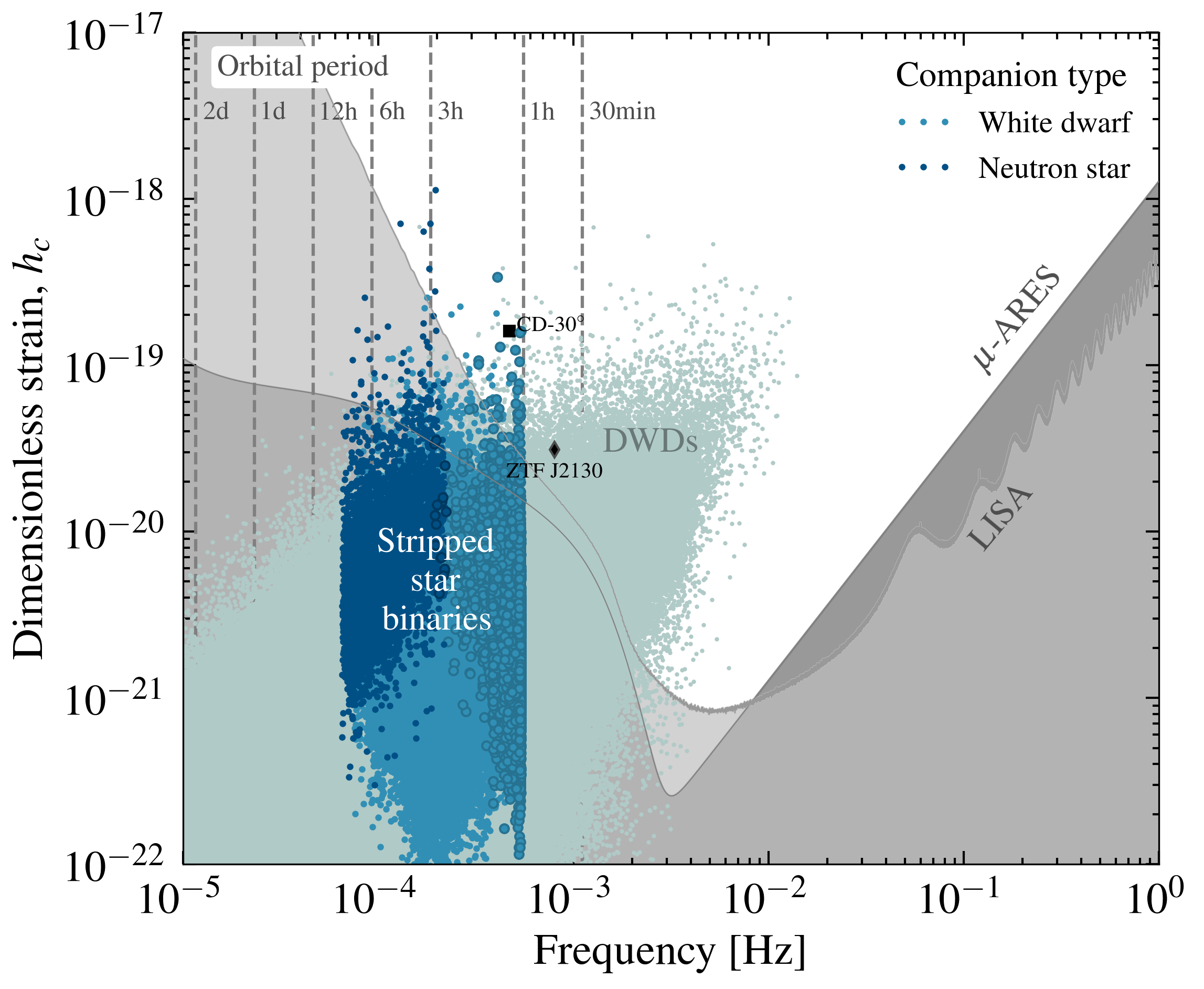}
\caption{The dimensionless, characteristic strain and gravitational wave frequency for stripped stars orbiting compact objects in one of our population realizations. Stripped stars with white dwarf and neutron star companions are marked with light and dark blue dots, respectively. The systems that are undergoing mass transfer are marked with a dark border surrounding the dot.
We show the sensitivity curves for LISA and $\mu$-ARES in light and dark gray backgrounds. 
Most of the stripped star binaries are undetectable by LISA, but the ones that are detectable are expected to be nearby ($\lesssim 1$ kpc) and can therefore be followed up in the electromagnetic regime. In this realization, 1 stripped star in orbit with a NS and 4 stripped stars orbiting WDs have SNR~$> 5$, assuming a 10 year operation time. Of these, 3 of the WD systems are interacting. 
For comparison, we also show a Monte Carlo model for double white dwarfs (labeled DWDs) in light green dots from \citet{2019MNRAS.490.5888L}. The detectable population of DWDs is characterized primarily by higher GW frequency, making them easier to detect with LISA, but they are also expected to be much fainter in the electromagnetic radiation than stripped star systems. 
The observed low-mass stripped stars orbiting white dwarfs, \CD\ and \ZJ, are marked with a black square and diamond respectively. We note that \ZJ\ is interacting, which likely explains why its orbit is smaller than what our model predicts (see \secref{sec:modeling}).
The dashed gray lines indicate different orbital periods. 
}
\label{fig:LISA_curve}
\end{figure*}

\subsubsection{Standard Model}\label{sec:standard}

We find that LISA will detect $3.1\pm 1.8$ stripped stars orbiting WDs and $0.1^{+0.4}_{-0.1}$ stripped stars orbiting NSs. 
About half of the detectable WD systems are expected to be mass transferring. Because of stochastic effects, we find realizations both with 0 and up to 2 stripped stars orbiting NSs and between 0 and 10 stripped stars orbiting WDs.

In \figref{fig:LISA_curve}, we display the results from one of the realizations. The figure shows the characteristic strain as a function of GW frequency for the Galactic population of stripped stars in tight orbit with compact objects, comparing to the sensitivity curve of LISA. This particular realization has one NS system and four WD systems with SNR~$>5$. Three of the WD systems are undergoing mass transfer. 

\begin{figure}
\centering
\includegraphics[width=\columnwidth]{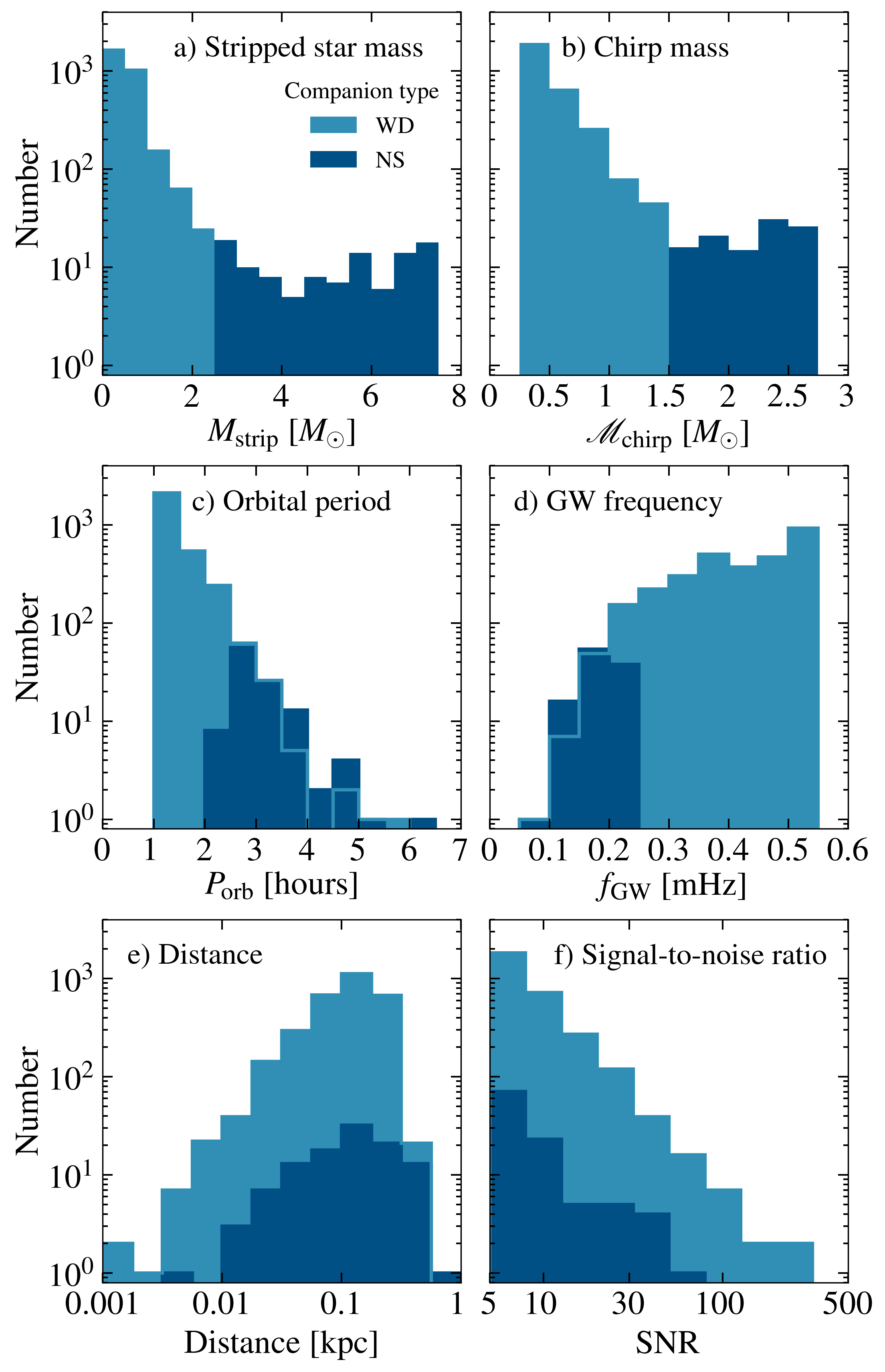}
\caption{Properties for the detectable population (SNR~$>5$) of stripped stars in tight orbit with compact objects. From top-left to bottom-right we show the mass distribution of the stripped stars, the chirp mass distribution, the orbital period distribution, the distribution of GW frequency, the distance, and the SNR distribution. The population of systems with NS companions are shown in dark blue and the population with WD companions are shown in light blue. We mark the contour of the WD distribution. 
To avoid low-number statistics, this figure is made using all detectable sources in the 1000 runs of our ``standard model''. }
\label{fig:prop}
\end{figure}

\figref{fig:LISA_curve} shows that the number of stripped stars in tight orbit with compact objects that are detectable by LISA constitutes only a small fraction of the entire population. The stripped star binaries pile up in a disadvantageous part of the LISA band between 0.06 and 0.6~mHz. In this frequency regime, the amplitude of a source needs to be more than an order of magnitude larger than for sources with $f_{\text{GW}}> 1$~mHz to be detectable with LISA (compare for instance with the double white dwarfs (DWDs) from \citealt{2019MNRAS.490.5888L} in \figref{fig:LISA_curve}). 
In practice, because stripped star binaries have similar chirp masses but lower frequencies compared to DWDs, LISA will only be able to detect them at shorter distances (see also \figref{fig:prop}e).
Other double compact object binaries are expected to overlap with the part of parameter space where stripped star binaries reside (see \tabref{tab:populations}). \citet{2019arXiv191207627S} predict about 6 double BHs detectable by LISA, while the Galactic double NS population is predicted to contain between 35-300 detectable systems \citep{2020ApJ...892L...9A, 2020MNRAS.492.3061L}.

We mark the location of the two detected, low-mass stripped star systems \CD\ and \ZJ\ in \figref{fig:LISA_curve}. 
\CD\ has a predicted SNR of $\sim 5$ for a 4 year observation time \citep{2018MNRAS.480..302K}. Its location in the diagram in \figref{fig:LISA_curve} is consistent with what we expect for detached systems containing a low-mass stripped star with a WD companion. 
\ZJ\ reaches the detection threshold of SNR$= 5$ after 10 years of the LISA mission. We note that it has a shorter orbital period (higher GW frequency) than allowed in our model. 
\ZJ\ is an interacting system where the stripped star is transferring material to its WD companion. This could explain the smaller radius of the stripped star in the system, which allows the orbit to be tighter \citep[cf.][]{2008AstL...34..620Y}. The short orbital period of \ZJ\ suggests that our treatment of mass transferring systems is not completely accurate and that there should be more detectable systems at higher GW frequencies. Therefore our predictions are likely to be underestimated. 

In addition to LISA, we also show the sensitivity curve of the recently proposed mission $\mu$-ARES \citep{2019arXiv190811391S} in \figref{fig:LISA_curve}. With its 400 million km long arms, $\mu$-ARES reaches the $\mu$Hz regime and would detect many more of the stripped star systems. Assuming an operation time of 10 years, we estimate that $\mu$-ARES would detect around 30 stripped star systems with SNR~$>5$, about 5 of which with NS companions. This is still a small fraction of the total Galactic population since $\mu$-ARES also will suffer from the astrophysical noise introduced by DWDs, seen as a bump in the sensitivity curve at the low-frequency end. However, since $\mu$-ARES reaches very low frequencies, the entire period range for stripped stars in tight orbit with compact objects can be probed. In addition, with the dozens of detectable systems, systematic studies of for example orbital tightening due to GW radiation and tides can be done.

To better understand the characteristics of the detectable population of stripped stars in tight orbit with compact objects, we display several of their properties in \figref{fig:prop}. To avoid stochastic effects, we show the results for all systems with SNR~$>5$ in the 1000 runs.

The mass distribution for detectable stripped stars (\figref{fig:prop}a) is skewed towards low-mass systems. This is both because low-mass stripped stars are more common and because they have smaller radii. Because they have smaller radii, they fit in tighter orbits and can reach higher GW frequencies, which are easier to detect with LISA. 
For NS systems, the mass distribution for detectable stripped stars is close to flat although the higher mass systems are rarer. The reason is that the GW amplitude depends on the chirp mass, meaning that a larger fraction of the higher mass systems will be detectable. The distribution of chirp masses of the detectable systems is shown in \figref{fig:prop}b and has a similar shape as the distribution of stripped star masses. The figure shows that of all detectable systems, the majority have WD companions and, therefore, have chirp masses below 1.5\Msun. 

The distribution of orbital periods of the detectable systems is shown in \figref{fig:prop}c. 
There is a sharp peak around 1-2 hours for stripped stars with WD companions and around 3 hours for stripped stars with NS companions. The main reason for these peaks is that LISA is more sensitive to short period binaries. 
For the systems including WDs, the orbits can be significantly tightened by GW radiation, causing a pile-up of systems at short orbital periods. 
The figure shows that it is unlikely that LISA detects a system with period longer than 5 hours, corresponding to a minimum GW frequency of around 0.1~mHz. \figref{fig:prop}d shows the distribution of gravitational wave frequencies. It is visible that most detectable WD systems have frequencies between $\sim 0.2-0.6$~mHz, while the detectable NS systems have $\sim 0.1-0.2$~mHz. The reason that most detectable systems have the described range of periods and GW frequencies is that LISA's sensitivity rapidly decreases with decreasing GW frequency. 

The GW amplitude depends on the distance to the source, leading the observable horizon for stripped star systems to be limited to $\sim 1$~kpc, as can be seen in \figref{fig:prop}e. The WD systems have somewhat lower masses (see \figref{fig:prop}b) and are therefore detectable within $\sim 500$~pc, while NS systems can be detected at larger distances. 

Most of the detectable systems are expected to have low SNR, as is shown in \figref{fig:prop}f. The figure shows that it is unlikely, but possible, that a source with SNR$\gtrsim 10$ will be detected.

\subsubsection{Model Variations}\label{sec:variations}

Aside from the ``standard model'', we also explore parameter variations, which reflect our lack of knowledge in different aspects. The exploration of these uncertainties covers many different axes of research, which are beyond the scope of this paper, but would merit their own exploration. 

For each parameter variation, we start from the standard set-up described in \secref{sec:modeling} and vary the uncertain parameters, one by one. For each parameter variation, we run the code 100 times and count the number of detectable systems each time.  Below, we describe how we vary the different parameters in each of these model variations.
\vspace{1ex}

\noindent \textbf{Wider orbits:}
In one model, we explore the impact of the uncertain post-common envelope period distribution. To do this, we assume a wider period distribution from $P_{\min}$ up to $10 \times P_{\min}$. In this case, orbital periods can be as long as 30 hours. 

\noindent \textbf{Efficient formation:}
In another model, we investigate how much the formation efficiency of a second stripped star affects the number of detectable systems. For this, we assume that no binary disruptions or mergers occur and all systems that create a first stripped star then also create a second stripped star in a tight orbit with a compact object, i.e., $f_{\text{eff}} = 100\%$. 

\noindent \textbf{Smaller radii:}
Shorter periods and thus higher GW frequencies than what we consider in the standard model are possible. As mentioned in \secref{sec:modeling}, the radii of stripped stars hardly change during their evolution, but these very small changes are sufficient to accommodate compact objects at even tighter orbits than what we previously have assumed. By investigating the evolutionary models of \citet{2018A&A...615A..78G}, we found that the radii of stripped stars can be up to 30\% smaller than what we assume during their central helium burning phase (steps 3 and 7 in \figref{fig:cartoon}).
Interestingly, this is sufficient to allow detached binaries with frequencies up to 0.9 mHz and thus also systems such as \ZJ\ are included \citep[cf.][]{2020ApJ...891...45K}. In this model variation, we therefore assume that stripped stars radii are 30\% smaller than those in the models of \citet{2018A&A...615A..78G}, resulting in orbital periods from 35 minutes to 5 hours. 

\noindent \textbf{Lower WD masses:}
In lack of carefully determined relations between stripped star mass and WD mass, we adopted a simplified approach that originates from evolutionary models of single stars (see \secref{sec:modeling}). Because envelope-stripping stops the growth of the stellar core, it is possible that we somewhat overestimated the WD masses. To explore the effect of the WD mass on the number of detectable systems, we create one model in which we set all WD progenitors more massive than 0.6\Msun\ to become WDs with masses of 0.6\Msun. The lower mass WDs have the same mass as their progenitors. 

\noindent \textbf{Higher star-formation rate:}
The Galactic star-formation rate is somewhat uncertain. To explore an optimistic scenario, we create a model in which we assume the Galactic star-formation rate is twice as high compared to what we previously assumed, i.e., 4\Msunyr \citep[e.g.,][]{2006Natur.439...45D}.

\noindent \textbf{Nominal operation time:}
Throughout this paper we assume a LISA mission duration of 10\,years, however this is yet to be confirmed and will be definitively set in early 2020s when ESA will fully adopt the mission. At present, the nominal mission lifetime is conservatively set to 4\,years. Therefore, we consider a model in which we assume the nominal operation time of 4 years. 
\vspace{1ex}

\begin{table}
\caption{The number of stripped stars orbiting white dwarfs ($N_{\text{WD}}$) and neutron stars ($N_{\text{NS}}$) that are detectable with LISA (SNR~$>5$). We show the numbers for different variations of relevant parameters (see \secref{sec:variations}).}
\label{tab:variations}
\begin{center}
\begin{tabular}{lccc}
\toprule \midrule
Model & $N_{\text{WD}}$ & $N_{\text{NS}}$ \\
\midrule
Standard (\secref{sec:standard}) & $3.1\pm 1.8$ & $0.1 ^{+0.4}_{-0.1}$ \\    
Wide orbits ($P_{\max} = 10P_{\min}$) & $0.6 ^{+0.8}_{-0.6}$ & $0.1^{+0.2}_{-0.1}$ \\  
Efficient formation ($f_{\text{eff}} = 100\%$) & $30 \pm 5$ & $1.1 \pm 1.0$ \\    
Smaller radii ($0.7R_{\star}$) & $100 \pm 11$ & $4.0 \pm 2.0$ \\  
Maximum WD mass 0.6\Msun & $2.9 \pm 1.7$ & $0.1^{+0.3}_{-0.1}$ \\    
Higher SFR (4\Msunyr) & $6.1 \pm 2.5$ & $0.3^{+0.6}_{-0.3}$ \\  
Nominal operation time (4\,yr) & $1.2 \pm 1.1$ & $0.0^{+0.2}_{-0.0}$ \\   
\bottomrule
\end{tabular}
\end{center}
\tablecomments{The settings of the ``standard model'' are: (1) period range between the minimum period and three times the minimum period, (2) $f_{\text{eff}} = 10\%$, meaning that 10\% of the systems that create a first stripped star also creates a second that orbits a compact object, (3) radii measured halfway through central helium burning in the models from \citet{2018A&A...615A..78G}, (4) masses of WDs mapped from initial masses following \citet{2018ApJ...866...21C} but with maximum mass of the progenitor mass of 1.3\Msun, (5) star-formation rate in the Milky Way of 2 \Msunyr, and (6) operation time of LISA of 10 years.}
\end{table}

The number of detectable stripped star systems in the different model variations are presented in \tabref{tab:variations}. As discussed in \secref{sec:standard}, about three systems are expected to have SNR~$>5$ in our ``standard model'' and that in about one out of ten cases one of them includes a NS.

\tabref{tab:variations} shows large differences in the number of detectable systems between the different model variations. 
The largest difference occurs when we assume that stripped stars are 30\% smaller. Then, the number of observable sources increases up to about 100 for WD systems and about 4 for NS systems. This suggests that LISA can provide strong constraints on the minimum radii of stripped stars.
The assumption for how efficient the formation of stripped stars orbiting compact objects also directly impacts the number of detectable systems. When assuming that $f_{\text{eff}} = 100\%$ instead of the standard $f_{\text{eff}} = 10\%$, the number of observable systems is an order of magnitude higher. 
The orbital period distribution also significantly affects the number of detectable systems. By increasing the maximum period from $3 \times P_{\min}$ to $10 \times P_{\min}$, the number of detectable systems decreases below one. 
We find that the other variations are less important for the number of detectable systems. 
When assuming that WDs cannot be more massive than 0.6\Msun, we find no significant difference in the number of detectable sources.
The detectable number is directly proportional to the assumed star-formation rate of the Milky Way. Therefore, in the run where we doubled the Galactic star-formation rate, the number of detectable systems is also doubled. 
The table also shows that maybe only one system is expected to be detected during the first 4 years of operation. 

The differences in the number of detectable systems between the variations are very interesting for constraining binary evolution models. If, for example, tens to hundreds of sources are detected by LISA, our ``standard model'' is inaccurate. If the sources are at higher frequencies than what we predict, stripped stars can be smaller than what we have assumed, meaning that the period distribution also will be different. The GW frequency of interacting systems will also help to constrain the conservative mass transfer. On the other hand, if LISA does not detect any more stripped star binaries than the already known \CD\ and \ZJ, it could suggest that the efficiency of common envelope ejection is higher than what we have assumed in this study. 

\citet{2020A&A...634A.126W} simulated the number of low-mass stripped stars (subdwarf B stars) in tight orbit with NSs using population synthesis and detailed evolutionary models \citep[see also][]{2018A&A...618A..14W}. They found that $\sim 100-300$ systems have SNR~$>1$ in the LISA band after 4 years of observations. In contrast, our standard model predicts only 30 systems will have SNR~$>1$ after 4 years of observations. Given that we consider a full range of possible masses, the discrepancy between the two predictions is puzzling. A combination of differences could provide explanations. \citet{2020A&A...634A.126W} assumed a higher Galactic star-formation rate of 5\Msunyr, 
they predicted a large Galactic population of sdB + NS systems created via common envelope ejection of at maximum 6000 systems, and, maybe the most important difference, they considered efficient common envelope ejections.

\section{Electromagnetic counterparts}\label{sec:EM}


\begin{table*}
\caption{Comparison of stripped star binaries with other monochromatic LISA sources. In the columns, we give in order: the type of source, the predicted number of systems that will be detected by LISA, the gravitational wave frequency, the range of chirp masses, and the typical distances out to which the sources are expected to be detected. (The distances are approximate, and we note that, for example, some double WDs in the Andromeda galaxy may be detectable \citep{2018ApJ...866L..20K}.)}
\label{tab:populations}
\begin{center}
\begin{tabular}{lcccc}
\toprule\midrule
GW source & Number & Frequency & Chirp mass & Distance \\
\midrule
Stripped star + NS$^a$ & $\sim 0-4^{\star}$ (SNR~$>5$) 
& $< 0.2$ mHz & $1.6-2.6$ \Msun & $\lesssim 1$ kpc \\

Stripped star + WD$^a$ & $\sim 0-100^{\star}$ (SNR~$>5$) 
& $< 0.6$ mHz & $0.3-1.6$ \Msun & $\lesssim 1$ kpc  \\

\midrule
Double WDs & $\gtrsim 10\,000^{b,c,\dagger}$ (SNR~$>7$) & $0.5-10$ mHz$^b$ & $0.1-1$\Msun$^b$ & $\lesssim 30$ kpc  \\ 



Double NSs & $\sim10 - 100 ^{\dagger}$ (SNR~$ > 7$)$^{d, e}$ & $0.1-10$ mHz$^{d}$ & $\sim 1.2\Msun$ & $\lesssim 140$ kpc \\ 

Double BHs$^f$ & $\sim 6^\star$ (SNR~$>7$) & $0.1-1$ mHz & $\sim3-50\Msun$ & $\lesssim 5000$ kpc \\ 

\bottomrule
\end{tabular}
\end{center}
\tablecomments{{$^\dagger$ $T_{\text{obs}} = 4$ years, $^\star$ $T_{\text{obs}} = 10 $ years. 
$^a$ This work, 
$^b$ \citet{2019MNRAS.490.5888L}, 
$^c$ \citet{2017MNRAS.470.1894K}, 
$^d$ \citet{2020MNRAS.492.3061L}, 
$^e$ \citet{2019arXiv191013436A}, 
$^f$ \citet{2019arXiv191207627S}.
}
}
\end{table*}

Other types of GW sources will be present in the predicted frequency range of stripped star binaries. In \tabref{tab:populations}, we provide the predicted numbers, GW frequencies, chirp masses and distances to sources that are expected to be monochromatic in the LISA band. The table shows that double compact objects occupy the same frequency range as stripped star binaries. The predicted rates of double compact objects are either similar (as in the case of double BHs and double NSs) or much larger (as in the case of double WDs) compared to stripped star binaries. Therefore, at frequencies $\lesssim 1$~mHz, it will be difficult to distinguish between stripped star binaries and double compact objects using the GW strain and frequency alone. 

However, there is a notable difference between stripped star binaries and double compact objects: stripped stars are bright electromagnetic sources. This makes them much easier to locate and study than many of the compact object binaries \citep[e.g.,][]{2019MNRAS.482.3656R}. For this reason, they are (1) important, low-frequency verification binaries for LISA, and (2) suitable for studying and constraining binary evolution, for example how double compact objects are created (e.g., \figref{fig:cartoon}). In this section, we discuss promising techniques for detecting stripped star binaries in the electromagnetic regime.

\begin{figure*}
\centering
\includegraphics[width=0.8\textwidth]{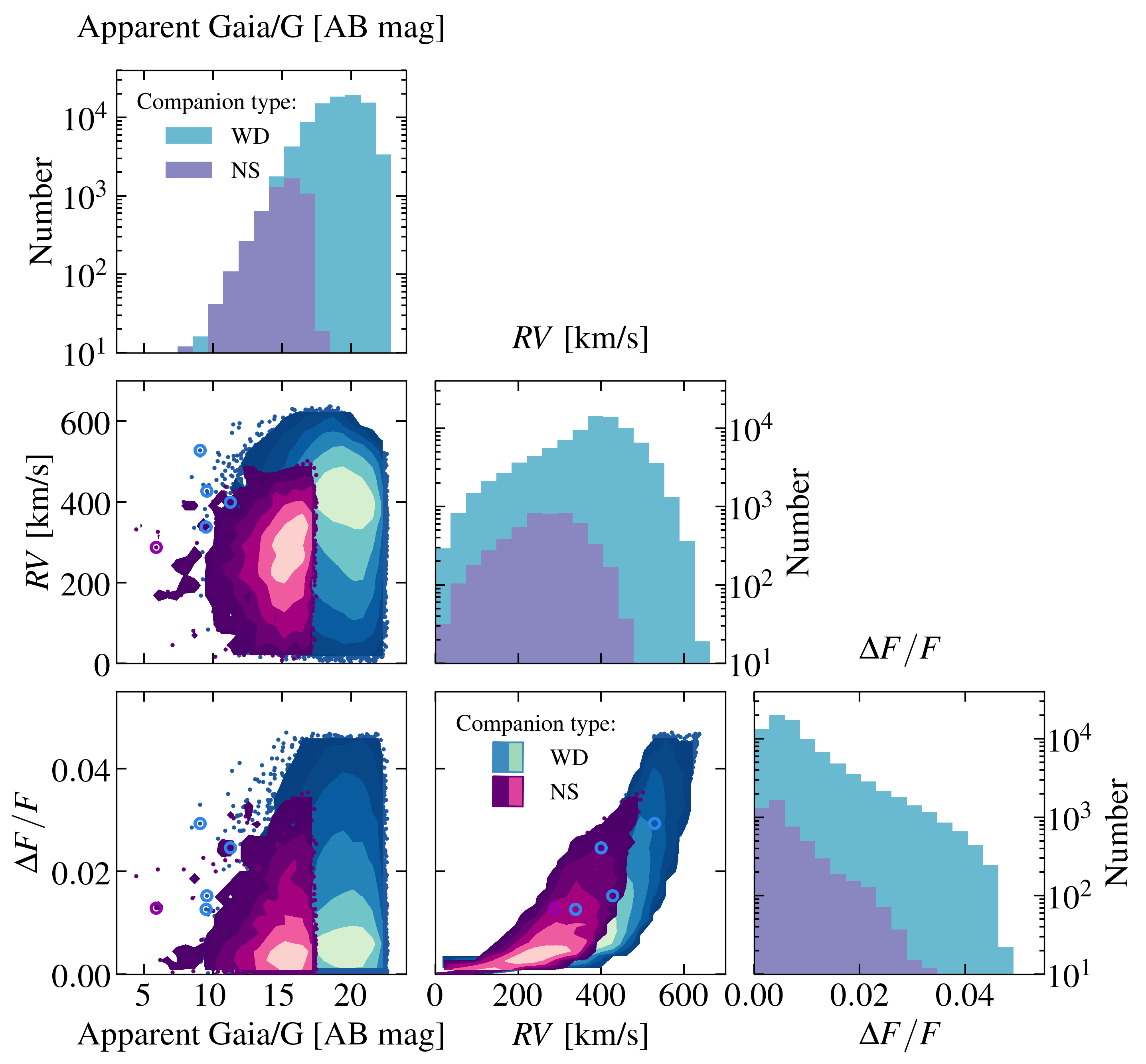}
\caption{Distributions and correlations of observable properties for the Galactic population of stripped stars in tight orbit with compact objects. In the diagonal, we show, from left to right, the apparent magnitude in the \textit{Gaia/G} band using the AB magnitude system and not accounting for extinction, the total radial velocity shift expected for the stripped star, and the fractional flux variations expected due to distortion of tidal forces over an orbital period. In the panels below the diagonal, we show the correlations between the different properties. We split the population into systems containing a WD and a NS and show the distributions for these two groups separately. The stripped star + WD population is shown in blue shades and the stripped star + NS population is shown in purple shades. From inside out, the contours enclose 38.3, 68.3, 86.6, 95.4, 98.8, and 99.7\% of the total population. 
The figure displays the same realization as in \figref{fig:LISA_curve}. We mark the systems that have SNR~$>5$ in LISA with a larger circle. The color of the circle is blue if the system contains a WD and purple if it contains a NS.}
\label{fig:triangle}
\end{figure*}

\subsection{Photometric Brightness of the Stripped Stars} 

Because compact objects are intrinsically faint (or do not emit light at all), a stripped star in orbit with a compact object should look much like an isolated stripped star:\ hot ($\gtrsim 30$~kK) and helium-rich \citep[with the exception of the lowest mass stripped stars, see][]{2018A&A...615A..78G}. Single stripped stars are expected to be rare as they should have been ejected from a binary system \citep{1994A&A...290..119P, 2019A&A...624A..66R}. Therefore are stripped stars that appear to be single good candidates for hosting compact companions. 

Stripped stars are the brightest in the short, extreme-ultraviolet ionizing wavelengths, but they are still relatively bright in the optical. The first histogram along the diagonal in \figref{fig:triangle} shows the distribution of apparent {\it Gaia}/$G$ magnitudes for the Galactic population of stripped stars in tight orbit with compact objects (using the AB magnitude system). The panel shows that the full population has apparent magnitudes between $\sim 10-25$ mag. This is significantly brighter than double WDs, whose magnitudes peak at $\sim 30-40$ mag \citep{2017MNRAS.470.1894K}. The stripped star binaries containing NSs are brighter than those containing WDs, reaching a maximum magnitude of $\sim 15$ mag. The reason is that we assume that the stripped stars accompanying NSs are more massive and therefore brighter than those accompanying WDs. Since the systems that are detectable by LISA are at $\sim 100-1\,000$~pc distance (see \figref{fig:prop}e), their apparent magnitudes are expected roughly between 3 and 15 mag. For the displayed model realization, the sources detectable by LISA are encircled in the middle and bottom left panels of \figref{fig:triangle}. The panels verify that the sources detectable by LISA are indeed brighter than the bulk of the population, simply because they are nearby. We note that we have not accounted for extinction, which significantly dims the distant stars, but is less severe for nearby stars. 

Despite their brightness, and even proximity in some cases, stripped stars have been difficult to detect in the Milky Way. One reason is that they are relatively young and therefore expected within the thin disk, at low Galactic latitudes, where extinction is high and stellar crowding is a known issue. When the stars are dimmed and reddened by extinction, hot stars are difficult to identify by their color, resulting in more sophisticated techniques being required \citep[see e.g.,][]{2020ApJ...891...45K}. The two known binaries
\CD\ and \ZJ\,, containing low-mass stripped stars and white dwarf companions, are indeed located at low Galactic latitudes \citep{2013A&A...554A..54G, 2020ApJ...891...45K}, which matches well with the evolutionary pathway that we are suggesting. With the large incoming data from {\it Gaia}, a large population of subdwarfs has now been identified \citep{2019A&A...621A..38G}. This database would be interesting to mine in the search for signatures of close compact companions.

\subsection{Orbital Motion}

Because of the very short orbital periods, we expect that the spectral lines of the stripped star show significant radial velocity (RV) variations if the star is in a tight orbit with a compact object. The second panel in the diagonal in \figref{fig:triangle} shows the distribution of the predicted total RV variation for the stripped star in each system, accounting for the viewing angle. About 90\% of the systems are expected to show RV variations higher than 200\kms\ and we predict RV variations up to 600\kms\ for WD systems and up to 450\kms\ for NS systems. The stripped stars accompanying a WD are expected to show larger variations compared to the ones accompanying a NS, because the low-mass stripped star systems have more equal mass ratio than the high-mass stripped star systems, where the stripped star is more massive than the NS.

Large surveys with multi-object spectrographs, such as 4MOST \citep{2014SPIE.9147E..0MD}, WEAVE \citep{2012SPIE.8446E..0PD}, and the Milky Way mapper survey included in SDSS-V \citep{2017arXiv171103234K}, will be excellent instruments for not only locating apparently single stripped stars but also identifying RV shifts.
Very nearby sources ($\lesssim 100$~pc) with somewhat longer orbital periods ($\gtrsim 3$ hours) and face-on inclinations could be detected as binary stars by their astrometric signature measured by the {\it Gaia} spacecraft \citep[$\sim 0.03$ mas,][]{2016A&A...595A...1G}\footnote{See \url{https://sci.esa.int/s/AG513LW}}. We expect that such very nearby systems are only a few. In the realization shown in \figref{fig:LISA_curve}, only one system is expected to have an astrometric signature that is detectable with {\it Gaia}.

\subsection{Deformation Leading to Variability}

For the tightest systems, the stripped star is deformed by tidal forces induced by the compact object, causing the star to take a drop-like shape. Because the visible surface area depends on the viewing angle, the brightness of the stripped star appears to increase and decrease twice every orbit for systems that are seen at, or close to, edge-on inclinations. 

We estimate this relative flux variation, also referred to as ellipsoidal modulation, for each system in the Galactic population of stripped stars binaries assuming that they are all tidally locked and following \citet[][see their Eq.~4]{2012MNRAS.422.2600B}. For the calculation, we assume that the linear limb-darkening coefficient is 0.2 and the gravity-darkening coefficient is 0.3, which is the best match for hot stars according to \citet{2017A&A...600A..30C}. The resulting distribution of expected ellipsoidal modulations are shown in the last panel along the diagonal in \figref{fig:triangle}. All systems in the population show small flux variations of less than 5\% for WD systems and less than 3\% for NS systems. This leads to magnitude variations of $\lesssim 0.05$ mag. 

The observed system \ZJ\ shows clear flux variations of up to 30\% \citep{2020ApJ...891...45K} and \CD\ has flux variations of about 10\% \citep{2013A&A...554A..54G}. Similar flux variations were also found by \citet{2017ApJ...851...28K} for another low-mass stripped star orbiting a WD (OW J074106.0–294811.0). These observed flux variations are significantly larger than our expectations. One reason could be that the darkening coefficients for the hot stars used as calibration by \citet{2017A&A...600A..30C} are different for stripped stars. We also note that \ZJ\ has a tighter orbit than the stars in our population, which affects the ellipsoidal modulations. In addition, \ZJ\ is eclipsed by an accretion disc, which further increases the system's flux variations. 
Because of the difference with observed systems, we consider our predicted flux variations to be lower limits. 

For detecting ellipsoidal modulations, {\it Gaia}, OGLE \citep{2015AcA....65....1U}, ZTF \citep{2019PASP..131a8002B}, and BlackGEM \citep{2016SPIE.9906E..64B} are suitable instruments since they have sufficient cadence and spatial resolution in the Galactic plane.

\subsection{X-ray Emission}\label{sec:Xrays}

Because the compact object orbits so close to the surface of the stripped star, it sweeps up material in the densest regions of the stellar wind, thus producing X-ray emission. During the accretion, the potential energy of the material falling onto the compact object is converted into luminosity following 
\begin{equation}\label{eq:LX}
    L_{X} = \varepsilon \dfrac{GM_{\text{CO}}\dot{M}_{\text{acc}}}{R_{\text{acc}}},
\end{equation}
where $M_{\text{CO}}$ is the mass of the compact object, $\dot{M}_{\text{acc}}$ is the mass accretion rate, $R_{\text{acc}}$ is the radius at which the accretion occurs, and $G$ is the gravitational constant. The parameter $\varepsilon$ describes how efficient the conversion to luminosity is. In the case of compact objects, the bolometric luminosity generated during accretion is primarily emitted in the X-rays, which is why we refer to it as $L_X$ in \eqref{eq:LX} \citep[see also][]{2009Sci...325.1222M,2013ApJ...764...41F}. 

We use \eqref{eq:LX} to estimate the X-ray luminosity of the Galactic stripped star binaries. For this, we assume that the accretion occurs on the surface of the compact object and assume that NSs have radii of 10 km, while we set the radii of WDs to follow the mass-radius relation of \citet{1997MNRAS.291..732T}. We set the parameter $\varepsilon$ to 1 in case the compact object accretes material at higher than 10\% of its Eddington accretion rate. When this is not the case, we compute $\varepsilon$ assuming an advection-dominated accretion flow following \citet{2012MNRAS.427.1580X}. The parameter $\varepsilon$ is between 0.01 and 1 for NS accretors, but can be several orders of magnitudes lower for WD accretors.

The assumed mass accretion rate is important for the final estimated X-ray luminosity. In the case of detached systems, we assume the Bondi-Hoyle accretion rate developed for stars sweeping up material from the surroundings \citep[][see also e.g., \citealt{2002apa..book.....F, 2008ApJS..174..223B, 2019ApJ...878L...4B}]{1944MNRAS.104..273B}. The material surrounding the compact object in the stripped star binaries is the stellar wind emitted from the stripped star. Since the winds of stripped stars are poorly understood, we need to make a few assumptions for the wind properties. First, we assume the wind mass-loss algorithm of \citet{2016A&A...593A.101K} for stripped stars less massive than 1.5\Msun\ and ten times lower values than what is predicted from the prescription of \citet{2000A&A...360..227N} for the higher mass stripped stars. We reduce the wind mass loss rates for the higher mass stripped stars to reach values consistent with the predictions from the theoretical wind models of \citet{2017A&A...607L...8V} for intermediate mass stripped stars. The resulting wind mass-loss range is between $10^{-20}$ and $10^{-7}\Msunyr$. We note that the low end of the mass-loss rate estimates is particularly uncertain since such low wind mass-loss rates have never been measured.
Second, we assume that the maximum wind speed, that is, the terminal wind speed, $v_{\infty}$, is 1.5 times larger than the escape speed from the stellar surface, which matches with the observed winds of WR stars \citep{1999isw..book.....L}. This results in terminal wind speeds of $v_{\infty} \sim 1500 - 2500$\kms, with the slower winds expected for the lower mass systems and vice versa. For the wind velocity profile, we assume a standard $\beta$-law \citep[$v(r) = v_{\infty} (1- R_{\mathrm{strip}}/r )^{\beta}$, where $r$ is the distance from the center of the stripped star and $R_{\mathrm{strip}}$ is the radius of the stripped star, see e.g.,][]{1979ApJS...39..481C} and set $\beta = 1$.

For systems that have initiated Roche-lobe overflow, we take a simplified approach and assume that all of the stellar wind is funnelled to the compact object. This means that the compact object accretes material at the same rate the stripped star is losing material. The mass transfer rate is likely higher during Roche-lobe overflow than what this assumption predicts, which could result in both higher accetion rates and higher X-ray luminosties \citep[see][]{2004MNRAS.349..181N, 2013ApJ...777..136W}. We note that when stripped stars have depleted helium in the center, they may expand and initiate interaction anew \citep{2020arXiv200301120L}. Since the star then swells rapidly, the mass transfer rate is expected to be much higher than what we assume for wind mass loss rate, which could lead the system to appear as an ultra-luminous X-ray source \citep[ULX, e.g.,][]{2019ApJ...886..118S}. However, because that phase is expected to be short-lasting, only a few such systems are expected \citep{2019ApJ...886..118S} and we, therefore, do not account for such systems. 

\begin{figure*}
\centering
\includegraphics[width=.7\textwidth]{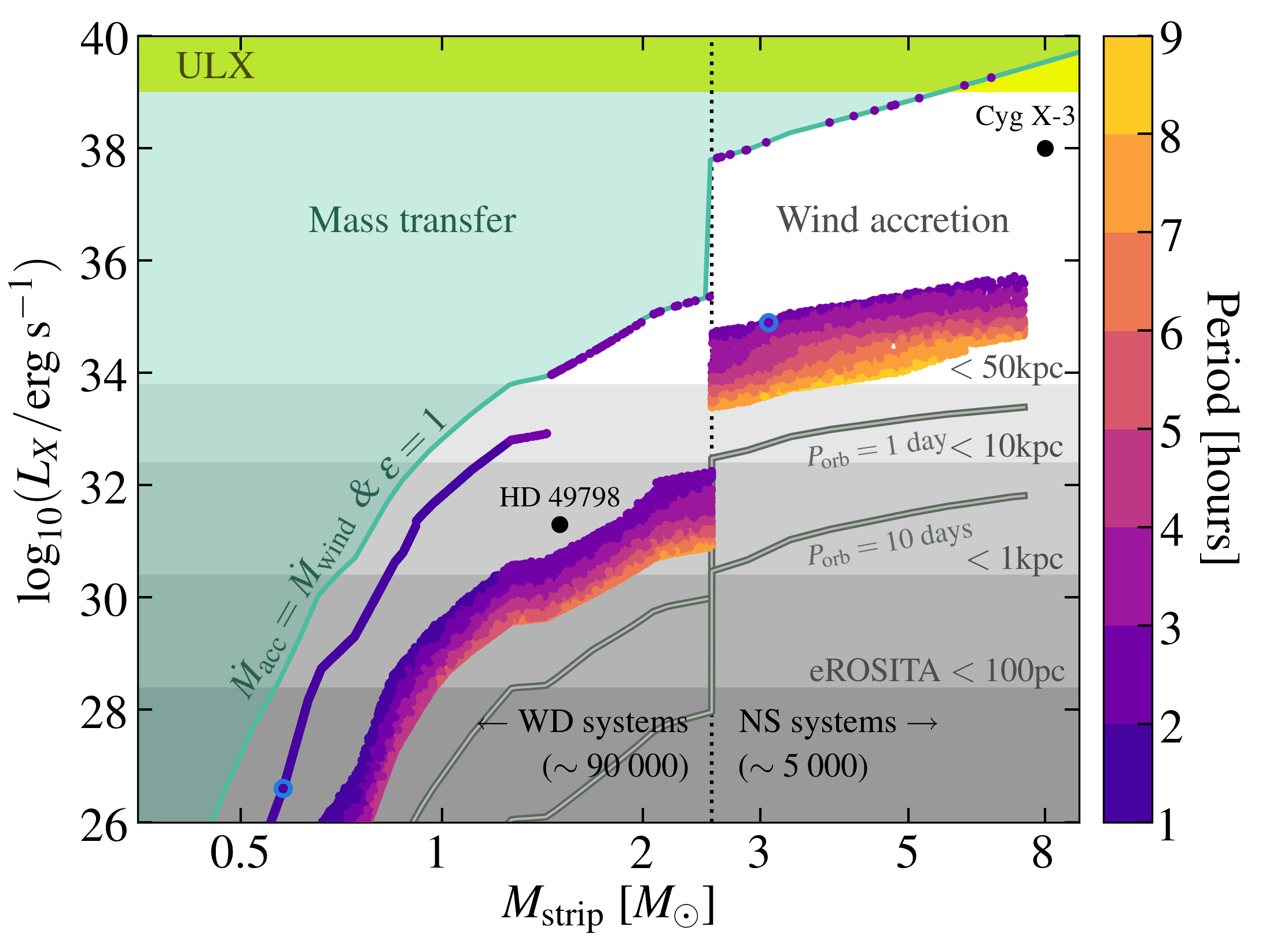}
\caption{Estimated X-ray luminosity of the Galactic population of stripped stars in tight orbit with white dwarfs and neutron stars as function of stripped star mass. Each system is represented by a dot that is colored according to the orbital period. 
The expected X-ray luminosity if the periods were 1 and 10 days are shown with a line and corresponding label.
For most systems, wind accretion onto the compact object is responsible for the X-ray emission. For the systems in which the stripped star is filling its Roche lobe, we assume that $\dot{M}_{\text{acc}} = \dot{M}_{\text{wind}}$, which increases their estimated X-ray luminosity by several orders of magnitudes. The efficiency parameter $\varepsilon$ is set to 1 for systems that exceed 10\% of the Eddington accretion rate, while for systems that accrete slower we adopt the algorithm of \citet{2012MNRAS.427.1580X} for advection dominated accretion-flows. 
We shade the region of $\dot{M}_{\text{acc}}>\dot{M}_{\text{wind}}$ and $\varepsilon = 1$ with green. The boundary of this region roughly corresponds to the distinction between mass-transferring and wind-fed systems. We note that the accretion rate can be higher during mass transfer compared to what we assume, meaning that some systems may be discovered inside the green-shaded region. 
The assumption for the companion type of the stripped stars is labeled and marked with a dotted line. It is possible that lower-mass stripped stars have neutron star companions \citep[cf.][]{2018A&A...618A..14W}, which could result in higher X-ray luminosities. One or a few Galactic systems could be so X-ray bright that they appear as ultraluminous X-ray sources \citep[see also][]{2019ApJ...886..118S}. 
We mark the 50, 10, 1 and 0.1 kpc horizons for eROSITA \citep[flux limit $\sim 2\times 10^{-14}$ erg cm$^{-2}$ s$^{-1}$,][]{2012arXiv1209.3114M} with gray shade and labels. 
This realization is the same as is shown in \figreftwo{fig:LISA_curve}{fig:triangle} and we encircle the binaries that have SNR$>5$ in LISA. We also mark the location of HD~49798, which is a 1.5\Msun\ stripped star likely orbiting a WD \citep{2009Sci...325.1222M}, and Cyg X-3, which is a $\sim 8 \Msun$ WR star orbiting a compact object \citep{1992Natur.355..703V}.}
\label{fig:LX}
\end{figure*}

In \figref{fig:LX}, we show the predicted X-ray luminosities as function of stripped star mass for the Galactic population of stripped stars in tight orbit with compact objects. We find that the X-ray luminosities of the stripped star systems can range from below $10^{26}\ergs$ up to $10^{40}\ergs$. In the case of NS accretors, the systems are typically detached and have $L_X \sim 10^{33} - 10^{36}\ergs$. That means that all the Galactic NS systems (about 5\,000 sources) would be within the flux detection limit of eROSITA \citep[$2 \times 10^{-14}$ erg cm$^{-2}$ s$^{-1}$,][]{2012arXiv1209.3114M} and about 75\% of them would be within the flux detection limit of ROSAT \citep[$5 \times 10^{-13}$ erg cm$^{-2}$ s$^{-1}$,][see also \citealt{2016A&A...588A.103B}]{1999A&A...349..389V}. 

The high fraction of X-ray bright sources from our simulated population suggests that stripped stars in tight orbits with compact objects may already exist in the ROSAT 2RXS catalog, which contains $\sim100\,000$ sources with unknown origin \citep{2016A&A...588A.103B}.
Rarely, we find X-ray luminosities in the regime of ULXs ($L_X \gtrsim 10^{39}\ergs$) and only one or a few such sources are expected.
Our predictions for the X-ray luminosity of NSs wind-fed by stripped stars are much lower compared to the observed luminosity of $\sim 10^{38}\ergs$ of the WR X-ray binary Cygnus X-3 \citep[Cyg X-3, e.g.,][]{2010MNRAS.402..767Z}. Cyg X-3 is an $\sim 8\Msun$ WR star and a $\gtrsim 3\Msun$ compact object on a 4.8 hour orbit \citep{1973A&A....25..387V, 1992Natur.355..703V}. Reasons for the high X-ray flux of Cyg X-3 compared to our predictions could be that the WR star has higher mass-loss rate or that accretion is more efficient than what our models predict. 

The expected X-ray luminosities from the WD systems is lower compared to what is expected from the NS systems. First, the apparent jump in X-ray luminosity visible in \figref{fig:LX} is due to the larger radii of WDs (see \eqref{eq:LX}). Second, the wind mass loss rate of stripped stars decreases with decreasing mass, which both affects the amount of material that is available for accretion and how efficiently it is accreted. With our simple assumptions, we predict that the WD systems have X-ray luminosities between $10^{28}\ergs$ and $10^{32}\ergs$ if they contain stripped stars more massive than around 1\Msun. This estimate matches relatively well with the observed X-ray luminosity of the system HD~49798, which is a 1.5\Msun\ stripped star that most likely orbits a WD in a 1.5 day orbit \citep[$L_X \sim 2 \times 10^{31}\ergs$,][]{2009Sci...325.1222M}. However, HD~49798 has a longer orbital period than what we consider, and at such long orbital periods we predict about two orders of magnitudes lower X-ray luminosities, which could imply that our assumption for the accretion efficiency is too low. 
Our estimates indicate that systems with lower-mass stripped stars ($\lesssim 1\Msun$) typically have very low X-ray luminosities ($\lesssim 10^{28}\ergs$), which makes them hard to detect in X-rays. 
However, if they are undergoing mass transfer, even systems containing the lowest mass stripped stars might become sufficiently X-ray bright to be detected. 
We estimate that only about 300 WD systems are sufficiently X-ray bright to be detectable with eROSITA. Most of these systems contain stripped stars with $\gtrsim 1\Msun$ and are located within a few kpc.  

We stress that our predictions for the X-ray luminosities are uncertain. 
For the higher-mass systems ($M_{\text{strip}} \gtrsim 1 \Msun$), the orbital period significantly affects the X-ray luminosities. If, for example, all systems would have orbital periods of 10~days the wind-fed systems at maximum reach X-ray luminosities of $10^{32}\ergs$, meaning that just a few percent would be within the flux detection limit of eROSITA. 
For the low-mass WD systems, both the conversion from mass accretion to X-ray luminosity ($\varepsilon$) and the assumed wind mass-loss rate of the stripped star have large impact on the X-ray luminosity. The theoretical wind mass-loss scheme of \citet{2016A&A...593A.101K} is verified with observed subdwarf stars, but no wind mass-loss rate lower than $10^{-11}\Msunyr$ could be measured, meaning that our predictions for stripped stars with lower mass than $\sim 1 \Msun$ are based on extrapolations. 
Apart from uncertainties related to the physical properties of the systems, the detectability of X-ray sources is also affected by crowding, since the binaries are expected in the Galactic plane, and extinction along the line of sight \citep[affecting especially energies $\lesssim 0.5$keV, see][]{2000ApJ...542..914W}. 
However, since the uncertainties are related to physical properties, we consider that characterizing the Galactic population of stripped star X-ray binaries appears to be a very promising technique to understand progenitors of double compact objects and to constrain uncertain properties such as the stellar winds of stripped stars. 
Since some of the expected LISA sources should have significant X-ray emission, we highlight that joint observations with LISA and Athena could be very valuable \citep{2020NatAs...4...26M}.

\section{Summary}\label{sec:summary}

We modeled the Galactic population of stars stripped of their hydrogen-envelopes that are in tight binaries with compact objects. These systems result from common envelope evolution and are believed to be a necessary step in the creation of merging double compact objects \citep[e.g.,][]{2017ApJ...846..170T}. Moreover, these systems are plausible progenitors for Ia supernovae \citep[e.g.,][]{2019ApJ...878..100W}. Because stripped star radii are small, common envelope evolution can tighten binary systems that contain such a star with a compact object to detectable GW frequencies. 

Since their emitted GWs are limited to the sub-mHz regime by the size of the stripped star ($\lesssim 0.6$~mHz), we expect that stripped star binaries will appear as monochromatic sources in the LISA band. They are good verification binary candidates in the low-frequency regime of LISA, since they are bright in electromagnetic radiation and can therefore be discovered and studied in advance of LISA's launch. 

We predict that the total Galactic population of stripped stars in tight orbit with compact objects is large: about 90\,000 WD systems and about 5\,000 NS systems. However, only a small subset of them will be detected by LISA. 
We estimate that $\sim 0-4$ stripped stars in orbit with a NS and $\sim 0-100$ stripped stars in orbit with a WD will be detectable by LISA with SNR~$>5$ over an observation time of 10 years. These systems are located within 1 kpc distance and have short orbital periods, typically between 1 and 5 hours. 
We provide a range for the number of detectable systems since the estimate is significantly affected by uncertainties primarily related to physical processes in interacting binaries and stellar properties. 
We do not expect any significant contribution from stripped star binaries to the astrophysical noise, because the double WD population is at least two orders of magnitude larger. 

Apart from some being LISA sources, stripped stars in tight orbits with compact objects are also excellent for constraining uncertain processes in binary evolution. The number of detectable systems is dependent on, for example, the outcome of common envelope evolution, the disruption of binaries related to the formation of a compact object, and the response of stripped stars to mass transfer \citep[cf.][]{2020A&A...634A.126W}. The actual number of detected LISA sources will thus provide valuable constraints.
In addition, individual observed systems will be useful for understanding for example wind mass loss from stripped stars, which currently is poorly understood but significantly impacts the future evolution of the binary \citep[e.g.,][]{2019MNRAS.486.4451G, 2020arXiv200301120L}.

With their relevance both as LISA verification binaries and for constraining binary evolution, it is important to study the Galactic population of stripped star binaries in advance of LISA's launch. 
We estimate that the Galactic population of stripped stars orbiting NSs have X-ray luminosities of typically $10^{33} - 10^{36}\ergs$ and could therefore be detectable by eROSITA. A promising technique to locate stripped star + NS systems could, therefore, be to identify X-ray sources and then follow them up with optical spectroscopy to verify that their optical spectrum appears as an isolated stripped star. If these targets are part of surveys with multi-object spectrographs such as 4MOST, WEAVE or SDSS-V, the binary orbit can be constrained by the detection of radial velocity variations. 
For the stripped stars orbiting WDs, the most promising detection technique could be to search for variability. If the system is sufficiently tight, the stripped star is distorted, which gives rise to ellipsoidal modulations that could be detectable with for example ZTF or OGLE if the system is viewed close to edge-on. Surveys with multi-object spectrographs will be ideal for detecting also stripped stars orbiting WDs. We expect that most of the stripped stars orbiting WDs are X-ray faint, but a fraction could be detectable with eROSITA. 
 
Although so far stripped stars in tight orbits with compact objects have been elusive even for state-of-the-art electromagnetic instruments, we show that they constitute a promising multi-messenger study case for the upcoming electromagnetic and gravitational wave facilities in 2030 and beyond.

\acknowledgments
%
The authors would like to thank Monica Colpi, Alberto Sesana, Silvia Toonen, Lieke van Son, Natasha Ivanova, Jim Fuller, Tony Piro, Eva Laplace, Selma de Mink, Stephen Justham, Santiago Torres-Rodr\'{i}guez, Chris Burns, Alex Ji, Kyle Kremer, Ilaria Caiazzo, Mathieu Renzo, Riley Connors, Guglielmo Mastroserio, and Manos Zapartas for fruitful discussions. 
YG acknowledges the funding from the Alvin E.\ Nashman fellowship for Theoretical Astrophysics. 
VK acknowledges support from the Netherlands Research
Council NWO (Rubicon 019.183EN.015 grant).
KB acknowledges funding from the Jeffery L. Bishop Fellowship.
AL acknowledges funding from the Observatoire de la C\^ote d'Azur and the Centre National de la Recherche Scientifique through the Programme National des Hautes Energies and the Programme National de Physique Stellaire.
MRD acknowledges support from the NSERC through grant RGPIN-2019-06186, the Canada Research Chairs Program, the Canadian Institute for Advanced Research (CIFAR), and the Dunlap Institute at the University of Toronto.
This research was supported in part by the National Science Foundation under Grant No.\ NSF PHY-1748958. 
Computing resources used for this work were made possible by a grant from the Ahmanson Foundation. 

\bibliographystyle{aasjournal}
\bibliography{references_bin.bib}

\end{document}